\documentclass[conference]{IEEEtran}

\usepackage{graphicx,multirow, array, rotating, booktabs, epstopdf, subfigure, balance,cite,soul,amsmath,amssymb,amstext,siunitx, tikz}

\graphicspath{{img/}}

\newcolumntype{L}[1]{>{\raggedright\let\newline\\\arraybackslash\hspace{0pt}}m{#1}}
\newcolumntype{C}[1]{>{\centering\let\newline\\\arraybackslash\hspace{0pt}}m{#1}}
\newcolumntype{R}[1]{>{\raggedleft\let\newline\\\arraybackslash\hspace{0pt}}m{#1}}

\usepackage[lined, algoruled, commentsnumbered]{algorithm2e}

\usepackage[normalem]{ulem}

\newcommand{\stkout}[1]{\ifmmode\text{\sout{\ensuremath{#1}}}\else\sout{#1}\fi}

\newcommand{\etal}{\textit{et al.}}

\usepackage[T1]{fontenc}

\usepackage{cprotect}
\usepackage[normalem]{ulem}
\usepackage[T1]{fontenc}

\usepackage{textcomp}
\usepackage{lipsum}

\newcommand\copyrighttext{%
	\footnotesize \textcopyright 2019 IEEE. Personal use of this material is permitted.
	Permission from IEEE must be obtained for all other uses, in any current or future
	media, including reprinting/republishing this material for advertising or promotional
	purposes, creating new collective works, for resale or redistribution to servers or
	lists, or reuse of any copyrighted component of this work in other works.}
\newcommand\copyrightnotice{%
	\begin{tikzpicture}[remember picture,overlay]
	\node[anchor=south,yshift=10pt] at (current page.south) {\fbox{\parbox{\dimexpr\textwidth-\fboxsep-\fboxrule\relax}{\copyrighttext}}};
	\end{tikzpicture}%
}

\begin{document}

\title{Synchronization for Diffusion-based Molecular Communication Systems via Faster Molecules}

\author{\IEEEauthorblockN{Mithun Mukherjee\IEEEauthorrefmark{1},   H.~Birkan Yilmaz\IEEEauthorrefmark{2}, Bishanka Brata Bhowmik\IEEEauthorrefmark{3}, Jaime Lloret\IEEEauthorrefmark{4}, and Yunrong Lv\IEEEauthorrefmark{1}}
	\IEEEauthorblockA{\IEEEauthorrefmark{1} Guangdong Provincial Key Lab of Petrochemical Equipment Fault Diagnosis,\\
		Guangdong University of Petrochemical Technology, China}
	\IEEEauthorblockA{\IEEEauthorrefmark{2} Department of Network Engineering, Polytechnic University of Catalonia, 08034 Barcelona, Spain}
	\IEEEauthorblockA{\IEEEauthorrefmark{3} Tripura University, India}
	\IEEEauthorblockA{\IEEEauthorrefmark{4} Universitat Politecnica de Valencia, Spain}
	Email: m.mukherjee@ieee.org, birkan.yilmaz@upc.edu, bishankabhowmik@tripurauniv.in, jlloret@dcom.upv.es}

\maketitle

\copyrightnotice

\begin{abstract}
In this paper, we address the symbol synchronization issue in molecular communication via diffusion (MCvD). Symbol synchronization among chemical sensors and nanomachines is one of the critical challenges to manage complex tasks in the nanonetworks with molecular communication (MC). As in diffusion-based MC, most of the molecules arrive at the receptor closer to the start of the symbol  duration, the wrong estimation of the  start of the symbol interval leads to high symbol detection error. By utilizing two types of molecules with different diffusion coefficients we propose a synchronization technique for MCvD. Moreover, we evaluate the symbol-error-rate performance under the proposed symbol synchronization scheme for equal and non-equal symbol duration in MCvD systems. 
\end{abstract}

\begin{IEEEkeywords}
Molecular communication, diffusion-based communication, synchronization, nanonetworks, receiver design
\end{IEEEkeywords}

%%%%%%%%%%%%%%%%%%%%%%%%%%%%%%%%%%%%%%%%%%%%%
%%%%%%%%%%%%%%%%%%%%%%%%%%%%%%%%%%%%%%%%%%%%%
%%%%%%%%%%%%%%%%%%%%%%%%%%%%%%%%%%%%%%%%%%%%%

\section{Introduction}
Molecular communication (MC), a bio-inspired approach, is emerging as a promising technique for communication in nanonetworks~\cite{NakanoBook2013, Farsad2016, AkyildizNanoNetworks2008}. MC that is foreseen as one of the driving technologies for Internet of Things at the biological and nanoscale domains can be widely applied to future wearable and implantable devices for healthcare, environmental protection, and nano-medicine applications~\cite{NakanoBook2013, Farsad2016, AkyildizNanoNetworks2008, BerezaMalcolm2014, Upadhyay2017,Jamali2017NanoBioSensor}.

Among many MC systems, we consider molecular communication via diffusion (MCvD) where the emitted molecules reach to the receiver through fluid medium via diffusion process. The MC systems have several distinct characteristics compared to traditional wireless and acoustic communications such as: 1) the signal is transmitted by changing the molecule releasing time, molecular concentration, and molecular type (i.e., different chemical structure), 2) since the generation rate of the bio-molecules depends on energy and chemical budgets~\cite{Alberts2014}, it is not always possible to maintain either fixed number of transmitted molecules in each symbol interval or fixed symbol duration in the transmitter, and 3) due to size and power consumption limitation, to manage complex tasks in nanonetworks, {\it synchronization} among several components, e.g., chemical sensors and molecular machines, is one of the major challenges in the nanonetworks.

In the MC systems, modulation of information is done in symbol slots, which are assumed to be synchronized for effective modulation and demodulation. Precise clock synchronization is an essential part of nanonetworks to perform collaborative tasks~\cite{ABADAL201174, Lin2016}. Furthermore, a clock synchronization pattern among molecular machines is suggested in~\cite{Moore2013}. Although the oscillation period is synchronized, the alignment of the clock is not adequately considered. A clock sequence is calculated in~\cite{BlindSyncShah2013} based on the molecular channel delay. The main limitation of this scheme is that the released frequency and clock offset are assumed to be fixed. To overcome this limitation, Jamali \etal{}~\cite{SchoberSync-ICC-2017} suggested two low-complexity symbol synchronization schemes, namely, peak observation-based and threshold-trigger scheme for the MCvD systems. The threshold-trigger synchronization is based on the idea that an increase of concentration of specific molecule triggers a response of a cell. Thus, instead of considering total symbol interval, the detection of information molecules is performed during the interval when the number of synchronization molecules is above the predefined threshold. Although the previous studies~\cite{SchoberSync-ICC-2017,BlindSyncShah2013,Lin2016} laid a strong background on symbol synchronization, the receiver complexity and same diffusion coefficient of the molecules still limit the performance in terms of throughput and number of transmitted molecules in MCvD. A small estimation error on the starting of the symbol duration based on the peak concentration level of synchronization molecules has adverse impact on the symbol detection since most of the molecules arrive at the start of the symbol interval.

%%%%%%%%%%%%%%%%%%%%%%%%%%%%%%%%%%%%%%%%%%%%%%%%%
\begin{figure*}[t!]%
	\centering
	\includegraphics[width=1.8\columnwidth]{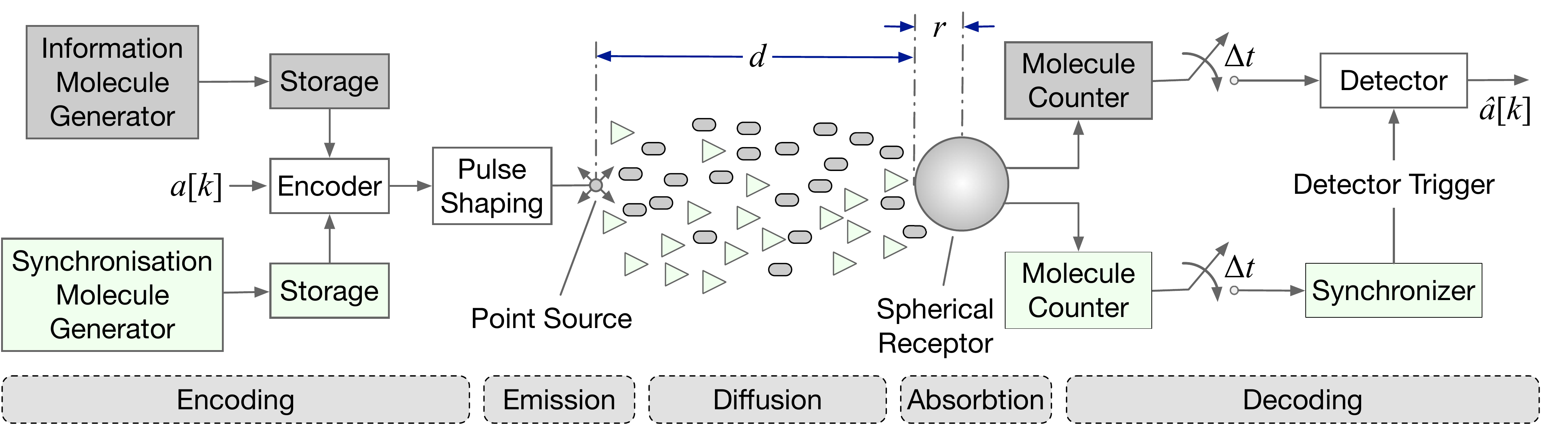}
	\caption{System model, processes, and the block diagram of an MCvD system with synchronization.}
	\label{fig:SystemModel}
		%\vspace{-0.15in}
\end{figure*}
%%%%%%%%%%%%%%%%%%%%%%%%%%%%%%%%%%%%%%%%%%%%%%%%%

%\subsection{Contributions} 
The main contributions of this paper include:

\begin{itemize}
\item We propose a symbol synchronization scheme for diffusion-based MC systems. We exploit the concentration of received molecules that heavily depends on diffusion coefficients. It is thus possible to estimate the starting of the symbol duration based on the high concentration level of synchronization molecules that are expected to arrive in advance due to higher diffusion coefficient than information molecules. 

\item The proposed synchronization method utilizes two types of molecules with different diffusion coefficients, therefore the proposed method does not require a training sequence to be sent.

\item In addition, since both of the synchronization and information molecules are released at the same time from the transmitter, the unequal symbol duration can be easily handled in the proposed per-symbol synchronization scheme for the MCvD systems.
\end{itemize}

The rest of the paper is organized as follows.  Section~\ref{sec:SystemModel} presents the MCvD system model.  The proposed frameworks for synchronization are discussed in Section~\ref{sec:ProposedScheme}. The simulation results are presented in Section~\ref{sec:SimulationResult}. Finally, conclusions are drawn in Section~\ref{sec:Conclusions}.

\section{System Model}\label{sec:SystemModel}
We consider a point-to-spherical MCvD system in a 3-dimensional (3-D) environment as illustrated in Fig.~\ref{fig:SystemModel}. The transmitter emits two types of molecules, type-$A$ molecule for information transmission and type-$B$ molecule for synchronization purpose. Assuming fluid propagation medium without any drift, the molecules that are released from the point source to the medium diffuse according to Brownian motion~\cite{BrownianEckford-2007}. 
We assume that the random displacement of two-types of molecules are independent to each other. Let $d$ be the distance between the point source and  the surface of the spherical receiver. The radius of the spherical receiver is denoted by $r$ and we assume fully absorbing boundary for the receiver where every molecule that collides with the surface is absorbed by the receiver and removed from the environment, thus, these molecules contribute to the received signal only once. 

Assume that intra-molecule collisions have negligible effect on the molecule's random movement. Using {\it first hitting} formula to model the {\it fully} absorbing reception process where the receiver is fully covered with perfect absorbing receptors, the hitting rate of type-$x$ molecules in 3-D environment by solving Fick's second law of diffusion is expressed as 
\begin{align}\label{eq:hittingfunction}
f_x(t)=\dfrac{r}{d+r}\dfrac{d}{\sqrt{4\pi D_xt^3}}\,\exp\!\left({-\dfrac{d^2}{4D_xt}}\right),
\end{align}
where $D_x$ is the diffusion coefficient for the type-$x$ molecule~\cite{Yilmaz3D-2014}. It is assumed that the diffusion coefficient is stationary throughout the medium. Since, (\ref{eq:hittingfunction}) has only one peak-value, the mean peak-time is obtained as follows
\begin{align}
\frac{\partial}{\partial t}\left(\dfrac{r}{d+r}\dfrac{d}{\sqrt{4\pi D_xt^3}}\,\exp\!\left({-\dfrac{d^2}{4D_xt}}\right)\right)=0.
\end{align}
By solving the above equation, we get $\mathbb{E}[t_{x,\textrm{peak}}]=\dfrac{d^2}{6D_x}.$

Furthermore, to obtain the fraction of molecules that hit the receiver until time $t$, we integrate $f_x(t)$ over the interval $t$ as 
\begin{align}
F_x(t)=\int_{0}^{t}f_x(\tau)\,\textrm{d}\tau = \dfrac{r}{d+r}\,\textrm{erfc}\left(\dfrac{d}{\sqrt{4 D_xt}}\right),
\end{align}
where $\textrm{erfc}(\cdot)$ is the complementary error function~\cite{Yilmaz3D-2014}.

%%%%%%%%%%%%%%%%%%%%%%%%%%%%%%%%%%%%%%%%%%%%%
%%%%%%%%%%%%%%%%%%%%%%%%%%%%%%%%%%%%%%%%%%%%%
%%%%%%%%%%%%%%%%%%%%%%%%%%%%%%%%%%%%%%%%%%%%%
\section{Receiver Structure and the Proposed Algorithm}\label{sec:ProposedScheme}
The receptors are deployed over a percentage of the receiver surface without any significant loss in the number of detected molecules~\cite{Akkaya2015}. In addition, the receiver can separately count two different types of molecules with different receptors. Thus, both synchronization and information carrying molecules can be counted separately via utilizing particular receptors. 

We denote the $k$-th data symbol by $a[k]$ and model $\{a[k]\}_{k=1}^{\infty}$ as a sequence of independent and identically distributed  variables that take on $\{1,0\}$ equiprobably for binary concentration shift keying (CSK)~\cite{MAHFUZ2010289}. Note that binary CSK is analogous to on-off keying in classical communications. Let $n_{\text{T},x}(k)$ be the number of released type-$x$ molecules for the $a[k]$-th symbol. Therefore, the hiting rate of received molecules for the $a[k]$-th symbol with type-$x$ at time $t$ becomes
\begin{align}
n_{\text{R},x}(k,t) & =  \underbrace{a[k]\,n_{\text{T},x}[k]\,f_x(t)}_{\textrm{desired signal}}\nonumber\\
& \quad +\underbrace{\sum_{m=1}^{k-1}a[m]\, n_{\text{T},x}[m]\,f_x(t-mT_s)}_{\textrm{intersymbol interference}}+\eta_b(t),
\end{align}
where $T_s$ is the symbol duration and $\eta_b(t)$ is the Brownian noise due to the random motion of the molecules. 
This non-stationary Brownian noise~\cite{BrownianNoiseShah2012} can be approximated by  Gaussian distribution $\mathcal{N}\big(0,N_{\text{T,x}}[k] \,F_x(t)(1-F_x(t))\big)$ by
considering Binomial arrival process $\mathcal{B}\left(N_{\text{T,x}}[k], F_x(t)\right)$~\cite{YilmazNoise2014}.

Finally, the total number of received molecules for $k$-th symbol until time $t$ is expressed as 
\begin{align}
N_{\text{R},x}(k,t) & =  a[k]\,N_{\text{T},x}[k]\,F_x(t)\nonumber\\
& \quad +\sum_{m=1}^{k-1}\,a[m]\,N_{\text{T},x}[m]\,F_x(t-mT_s)+N_b,
\end{align}
where $N_{\text{T},x}[k]$ and $N_b=\int_{0}^{t}\eta_b(t)\textrm{d}t$ are the total number of released type-$x$ molecules for the $k$-th symbol and total number of noise molecules due to random molecular motion, respectively.
Thereafter, we define the signal-to-noise ratio for type-$x$ molecules as $\text{SNR}_x=\mathbb{E}[n_{\text{T},x,k}(t)\,f_x(t)]/\mathbb{E}[\eta_b(t)]$. 

%%%%%%%%%%%%%%%%%%%%%%%%%%%%%%%%%%%%%%%%%%%%%%%%%
\begin{figure}[t!]%
	\centering
	\includegraphics[width=0.9\columnwidth]{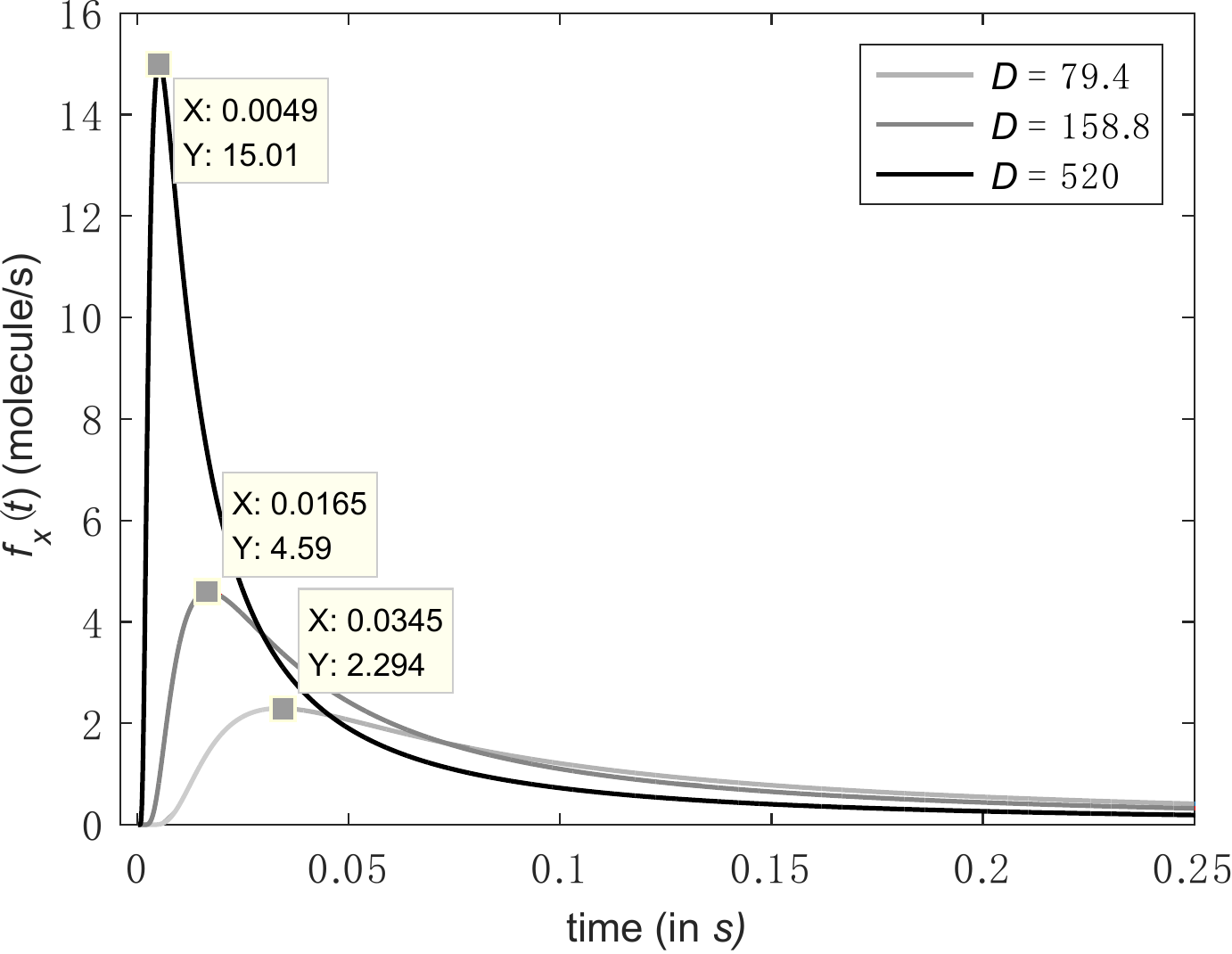}
	\caption{Hitting rate $f(t)$ for $r=\SI{4}{\micro\meter}$, $d=\SI{4}{\micro\meter}$, with different diffusion coefficient $D_x$. Note that $\mathbb{E}[t_{x,\textrm{peak}}]=\SI{0.0345}{\second}, \SI{0.0165}{\second},$ and $\SI{0.0049}{\second}$ for $D_x=\SI{79.4}{\micro\meter\squared\per\second}, \SI{158.8}{\micro\meter\squared\per\second},$ and $\SI{520}{\micro\meter\squared\per\second}$, respectively.}
	\label{fig:HittingRate}
	%	%\vspace{-0.12in}
\end{figure}
%%%%%%%%%%%%%%%%%%%%%%%%%%%%%%%%%%%%%%%%%%%%%%%%%

{\it Symbol detection with the proposed synchronization:}
The proposed symbol synchronization scheme in MCvD systems has the following two aspects:
\begin{enumerate}
	\item As observed in Fig.~\ref{fig:HittingRate}, the peak of received molecule  ratio for higher diffusion coefficient is obtained earlier than the molecules with low diffusion coefficient. To exploiting this feature in the MCvD systems, the diffusion coefficient of the synchronization molecules is assumed to be higher than that of the information molecules. Therefore, based on the high concentration level of the received synchronization molecules, which is earlier than the information molecules,  the receptors for information molecules estimate the starting of the symbol duration, thereafter, start to count the information molecules for the $k$-th symbol. 
	\item Since we consider per-symbol synchronization, the proposed technique can be used for the non-equal symbol duration and emitting-frequency. In addition, it also relaxes the constraint of maintaining a fixed clock-frequency at the receiver. 
\end{enumerate}

%%%%%%%%%%%%%%%%%%%%%%%%%%%%%%%%%%%%%%%%%%%%%
\begin{algorithm}[t]
	\label{Algorithm1}
	\footnotesize
	\SetAlgoLined
	\nl \Begin{		
		\nl	\For{$k=1$ to $K$}{
			\nl Estimate $\hat{t}_{\textrm{sync, peak}}(k)$ for synchronization molecules related for the $k$-th information symbol\;
			\nl $\hat{t}_{\text{info,start}}(k) = \hat{t}_{\textrm{sync, peak}}(k)$\;
			\nl $\hat{T}(k) = \hat{t}_{\textrm{info, start}}(k+1)-\hat{t}_{\textrm{info, start}}(k)$ \tcc{Estimated $k$-th information symbol}
				\nl\uIf{$N_{R,\text{info}}(k,T(k))>\text{Th}$}{
					\nl $a[k]=1$	
				}
				\Else{\nl $a[k]=0$\;}	
	}}
	\caption{Proposed synchronization scheme}
\end{algorithm}
%%%%%%%%%%%%%%%%%%%%%%%%%%%%%%%%%%%%%%%%%%%%%

Algorithm~\ref{Algorithm1} presents the main steps for our proposed schemes, where  $\hat{t}_{\text{sync, peak}}(k)$ is the estimated peak time of the received synchronization molecule for the $k$-th symbol, $\hat{t}_{\text{info,start}}(k)$ is the estimated start time for the $k$-th information symbol, and $K$ is the total number of information symbols. Basically, the transmitter emits the synchronization molecules on a {\it per-symbol} basis, the receiver takes the symbol duration information from the peak of the current synchronization molecules until it detects the next peak of the synchronization molecules. Although the selection of decision threshold $Th$ is itself a novel work, addressing these issues is beyond the scope of this paper. We utilize binary CSK, and we use a fixed threshold simply as $Th={N_{T,\textrm{info}}[k]}/{2}$.

\begin{figure}[!t]
	\centering
	\subfigure[a][CSK without synchronization.]{\label{fig:1} \includegraphics[width=0.98\columnwidth]{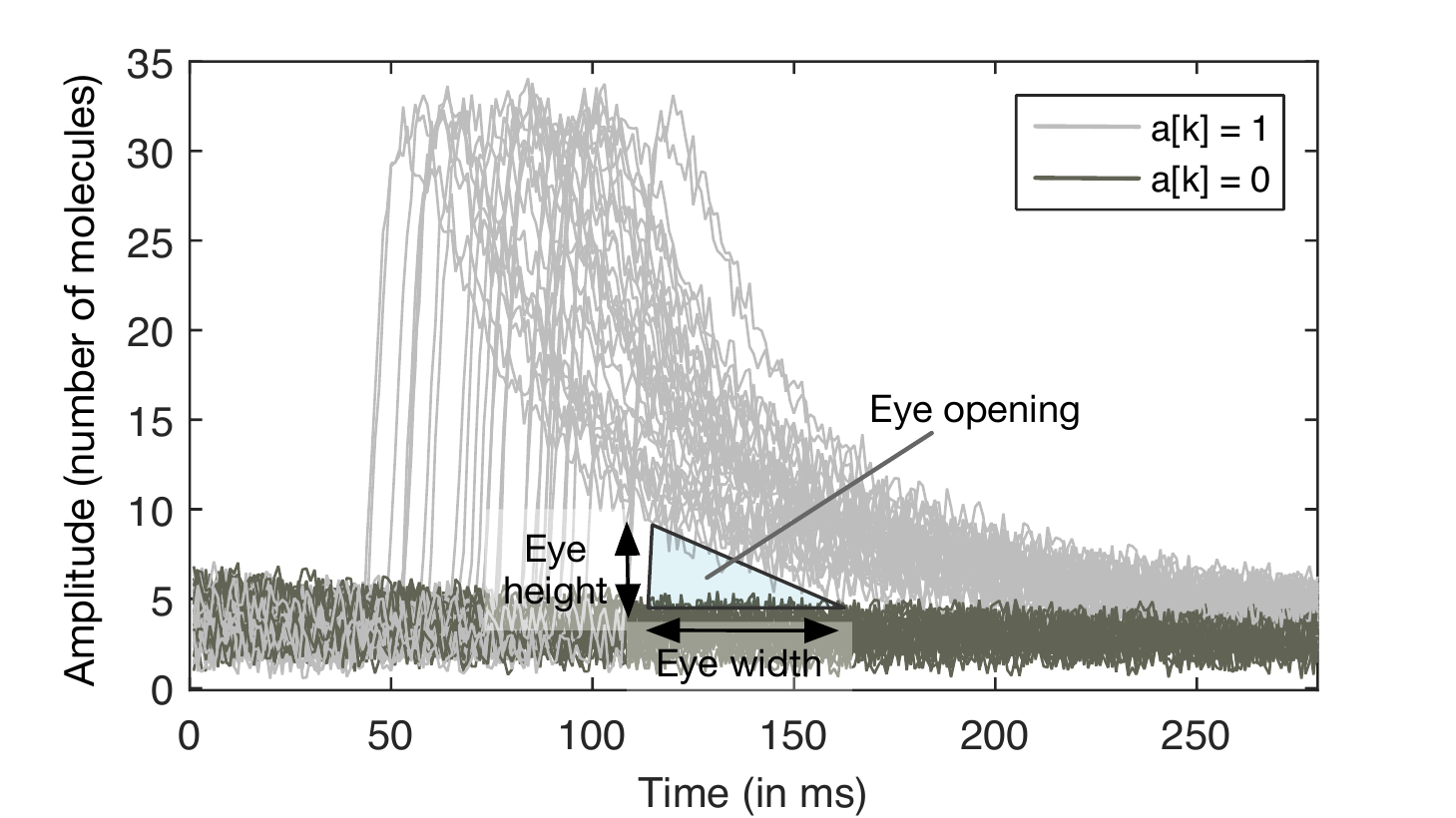}}\\
	%\vspace{-4pt}
	\subfigure[b][Proposed scheme with synchronization.]{\label{fig:2} \includegraphics[width=0.98\columnwidth]{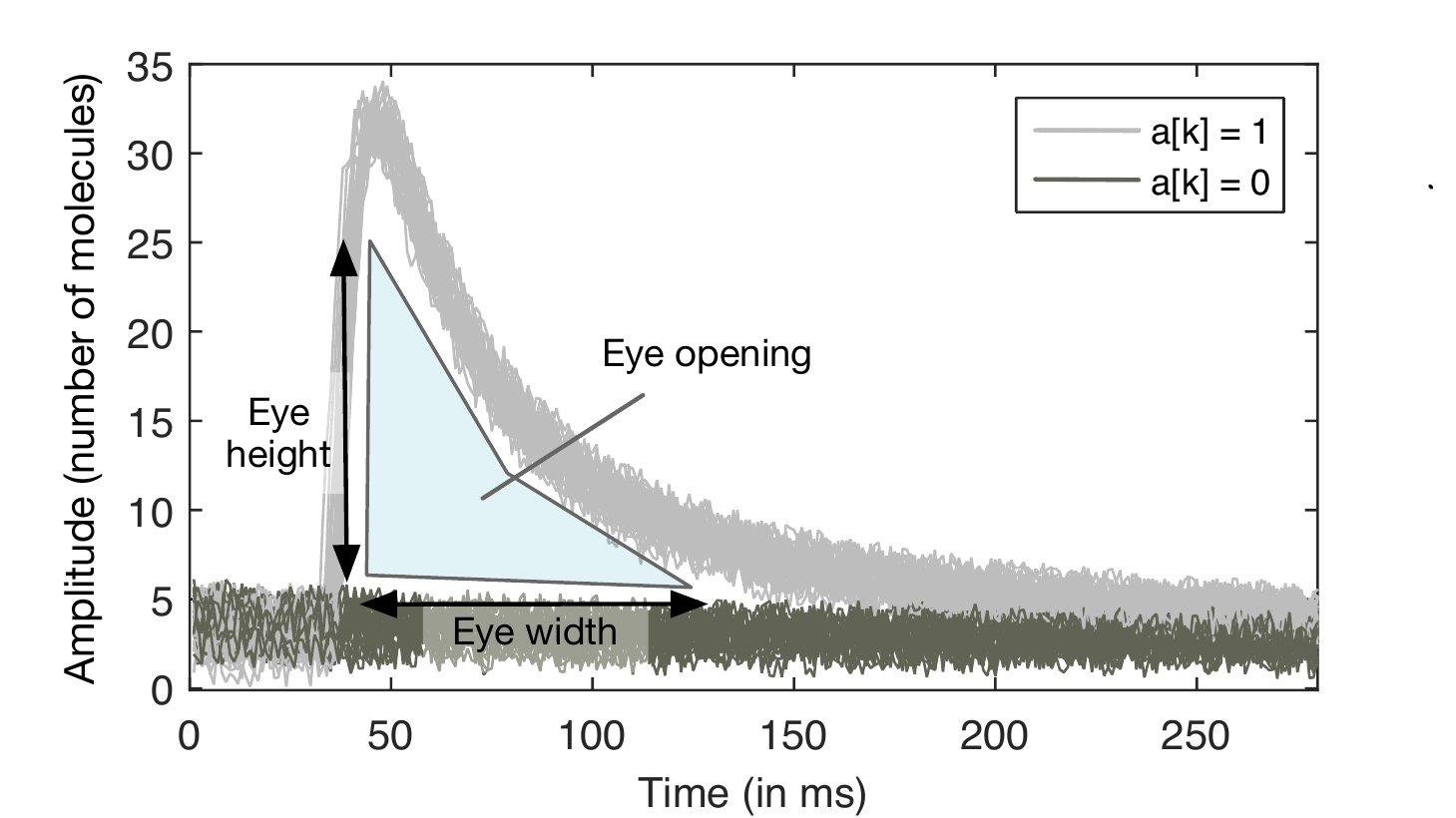}}
	\caption{Eye-diagram of the received information molecules.  We consider the non-equal symbol duration for the emitted molecules. The  starting of the each symbol is estimated, thereafter all the received information molecules are plotted in an overlapping manner. Simulation parameters are given in Table~\ref{Table:SimulationParameter}.}
	\label{fig:EyeDiagram}
	%	%\vspace{-0.1in}
\end{figure}

{\it Non-equal symbol duration:} Let $T(k)=(1+\psi)T_s$ be the symbol duration of the $k$-th time slot for information and synchronization molecules, where $-0.5\!<\!\psi\!<\!0.5$  is a truncated Gaussian random variable with zero mean and variance $\sigma_{\text{symbol}}^2$. If the transmitter lacks of information molecules due to the variation of energy and chemical budgets then non-equal symbol duration can be considered, otherwise $\psi=0$. Although storage element is assumed in the transmitter to supply molecules, the release time may change due to several physical behaviors. Therefore, the starting time as well as symbol duration of the release molecules are not same for all transmitted symbols. To highlight the impact of the variation in release time for the emitted molecules, Fig.~\ref{fig:EyeDiagram} illustrates the eye-diagram for the received information molecules assuming non-equal symbol duration. 
It is observed that the eye-height is higher in the proposed scheme compared to the binary CSK without synchronization. The eye-height determines the eye closure due to molecular noise. As the eye-width corresponds to the time interval of the received molecules, the wider eye-opening results less inter-symbol interference. As observed in Fig.~\ref{fig:EyeDiagram}, eye-width is more wider in the proposed scheme compared to the CSK without synchronization. Besides, these metrics are useful to measure the standard deviation of the received molecules.
Therefore, it is clearly observed that the proposed symbol synchronization scheme effectively handles the variation in release time by properly estimating the peak of the synchronization molecules compared to the binary CSK without synchronization. Moreover, it is easy to implement the proposed low-complexity symbol synchronization scheme in the small devices without  significantly increasing the hardware complexity at the receiver. The interested reader may refer to~\cite{MithunANTS18} for per-block synchronization in MCvD systems.

%%%%%%%%%%%%%%%%%%%%%%%%%%%%%%%%%%%%%%%%%%%%%%%%%
%%%%%%%%%%%%%%%%%%%%%%%%%%%%%%%%%%%%%%%%%%%%%%%%%

%%%%%%%%%%%%%%%%%%%%%%%%%%%%%%%%%%%%%%%%%%%%%
%%%%%%%%%%%%%%%%%%%%%%%%%%%%%%%%%%%%%%%%%%%%%
%%%%%%%%%%%%%%%%%%%%%%%%%%%%%%%%%%%%%%%%%%%%%
\section{Simulation Results}\label{sec:SimulationResult}
%%%%%%%%%%%%%%%%%%%%%%%%%%%%%%%%%%%%%%%%%%%%%%%%%
\begin{table}[!t]
	\centering
	\caption{Simulation Parameters}
	\label{Table:SimulationParameter}
	\footnotesize
	\renewcommand{\arraystretch}{1.1}
	%\vspace*{-\baselineskip}
	\begin{tabular}{R{0.45in}L{1.8in}L{0.65in}}\toprule
		Parameter & Definition & Value \\\midrule
		$D_A$ & Diffusion coefficient of type-A (information) molecule & \SI{79.4}{\micro\meter\squared\per\second}\\
		$D_B$ & Diffusion coefficient of type-B (synchronization) molecule & \SI{158.8}{\micro\meter\squared\per\second}\\
		$r$ & Radius of the spherical receiver & \SI{2}{\micro\meter}\\
		$d$ & Distance between transmitter and surface of the corresponding receiver & \SI{4}{\micro\meter}\\
		$T_s$ & Fixed symbol duration & \SI{380}{\milli\second}\\
		$\sigma_{\text{symbol}}^2$ & Variance of symbol duration & $0 \sim 0.3$\\
		$\Delta t$ & Sampling time at the receiver & \SI{10}{\micro\second}\\
		\bottomrule
	\end{tabular}
\end{table}
%%%%%%%%%%%%%%%%%%%%%%%%%%%%%%%%%%%%%%%%%%%%%%%%%
To compare with traditional binary CSK~\cite{MAHFUZ2010289} for the MCvD systems, we consider the normalized synchronization error as~\cite{SchoberSync-ICC-2017} $\bar{e}= \mathbb{E}[e(k)]/\mathbb{E}[T(k)]$, where the synchronization error is defined as $e(k) = |\hat{t}_{\text{sync, peak}}(k) -{t}_{\text{sync, peak}}(k)|$, where ${t}_{\text{sync, peak}}(k)$ is the actual peak time of the received synchronization molecules for the $k$-th symbol. When the synchronization scheme fails to detect the actual peak of the synchronization molecules, then the starting of symbol duration becomes erroneous, resulting wrong estimation of the detected information molecule count. 

The simulation parameters are summarized in Table \ref{Table:SimulationParameter}. The results are averaged over 100 different runs with at least $10^5$ symbols in each run. For each synchronization and information bearing symbols, we use 1000  molecules.

%%%%%%%%%%%%%%%%%%%%%%%
\begin{figure}[t!]%
	\centering
	\includegraphics[width=0.97\columnwidth]{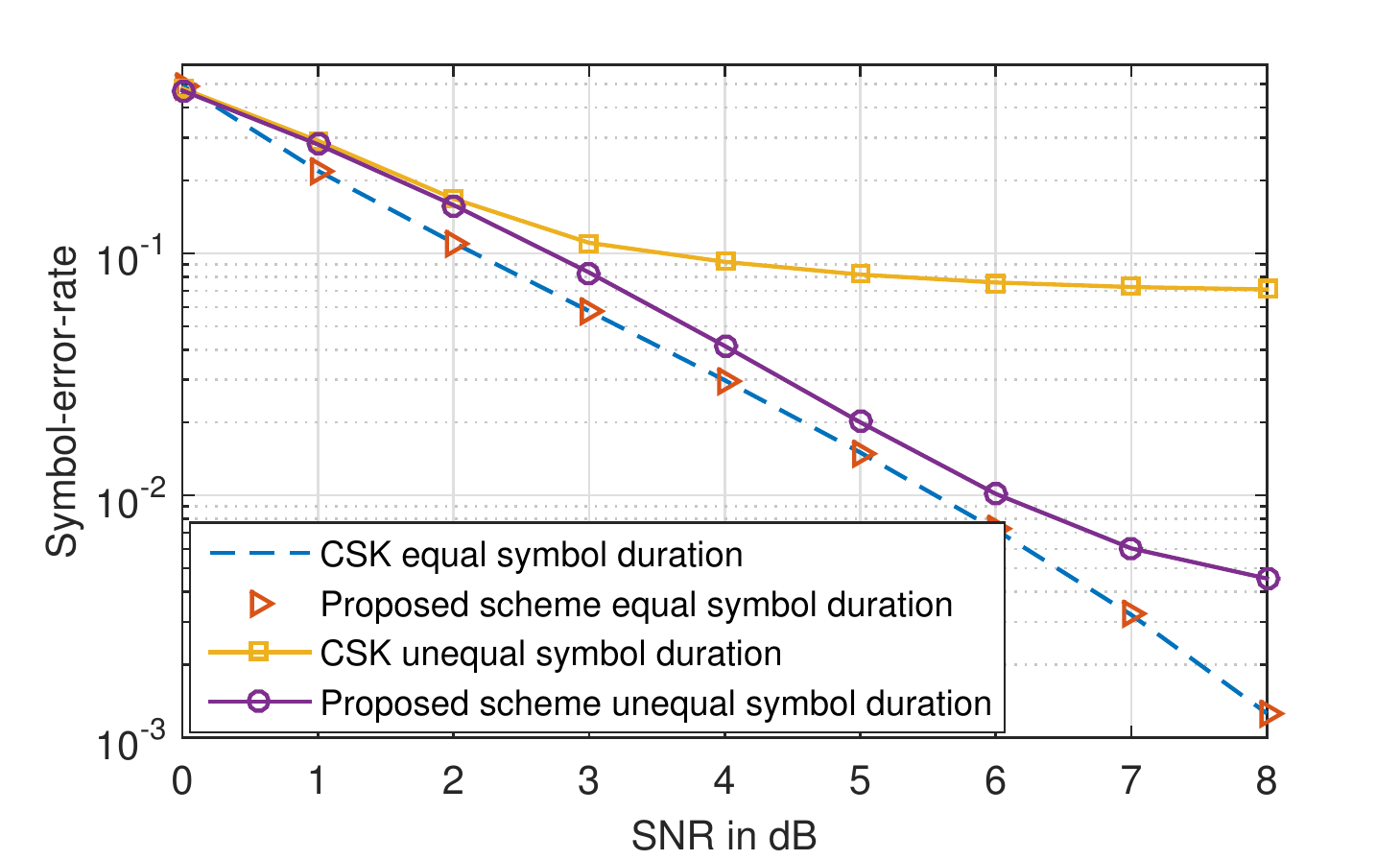}
	\caption{SER performance with various SNR values with perfect synchronization. We consider $\sigma_{\text{symbol}}^2=0.1$ for unequal symbol duration.}
	\label{fig:BERvsSNR}
\end{figure}
Fig.~\ref{fig:BERvsSNR} illustrates the symbol-error-rate (SER) performance with different SNR values. We use the same SNR values for both synchronization and information molecules. In fact, SER performance is same for the proposed scheme and traditional CSK for equal symbol duration without any synchronization error. However, proposed scheme performs significantly better than CSK when the symbol duration varies. A reduced SER is obtained in the proposed synchronization scheme compared to traditional CSK. Moreover, the error-floor, where the SER performance curve does not decrease significantly with the increase of the SNR, reaches earlier in the traditional CSK compared to the proposed scheme for the MCvD systems.

\begin{figure}[t!]%
	\centering
	\includegraphics[width=0.97\columnwidth]{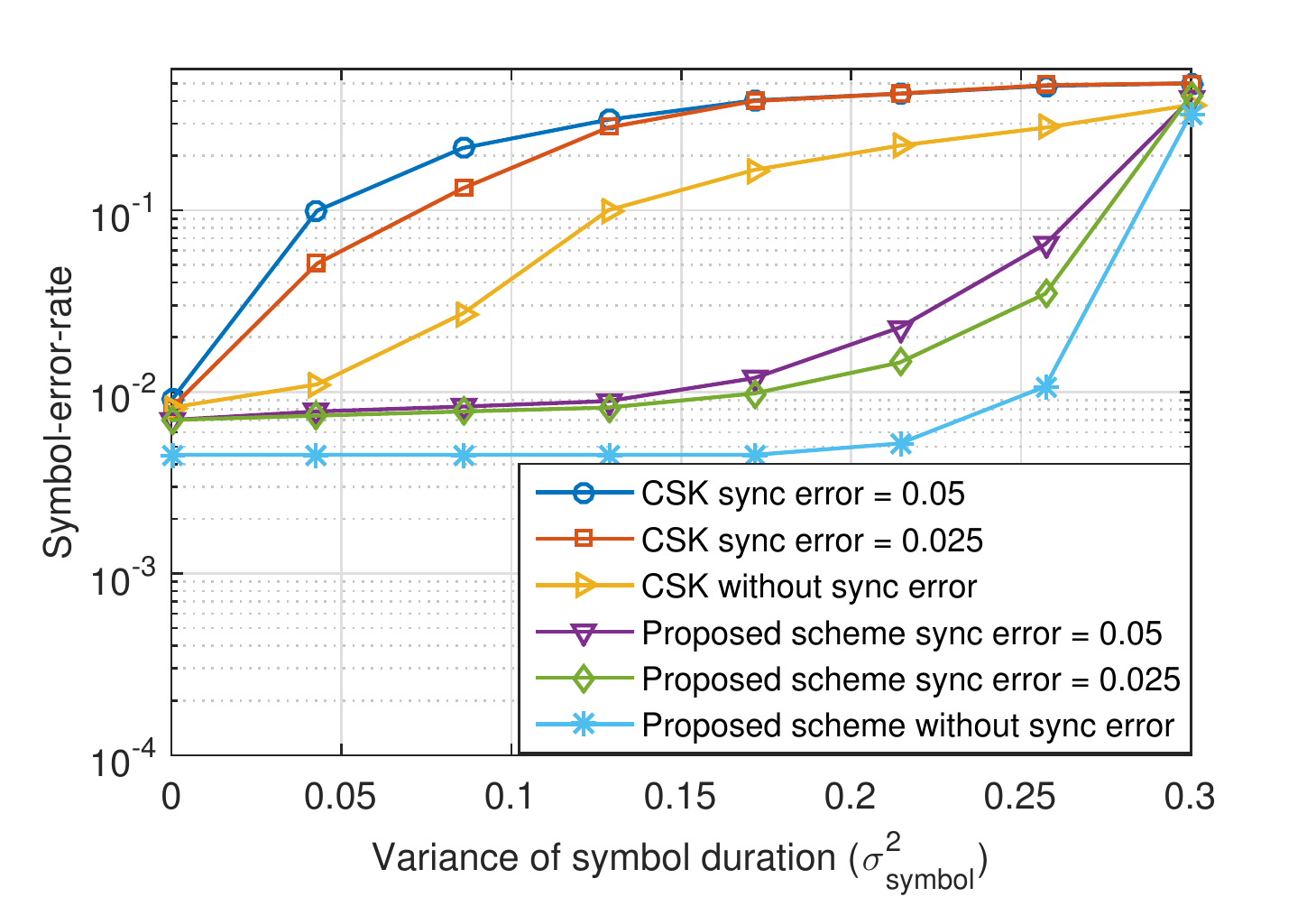}
	\caption{SER performance with the variance of symbol duration at $\SI{8}{\decibel}$ SNR for both synchronization and information bearing molecules.}
	\label{fig:SymbolVariation}
		\vspace{-0.14in}
\end{figure}

The effect of unequal symbol duration is shown in Fig.~\ref{fig:SymbolVariation}. The proposed scheme outperforms binary CSK in presence of unequal symbol duration even with a small value of its variance, e.g., \mbox{$\sigma_{\text{symbol}}^2 \simeq 0.1$}. Furthermore, we obtain the \mbox{SER  $\simeq 1\times10^{-2}$} with proposed synchronization at $\sigma_{\text{symbol}}^2 = 0.2$, whereas most of the symbol are erroneous in traditional binary CSK for the MCvD systems. These results show that the proposed synchronization scheme handles the unequal symbol duration by effective estimation of the peak of the synchronization molecules, thereafter, the starting of symbol duration for the information molecules. 
%%%%%%%%%%%%%%%%%%%%%%

\begin{figure}[t!]%
	\centering
	\includegraphics[width=0.97\columnwidth]{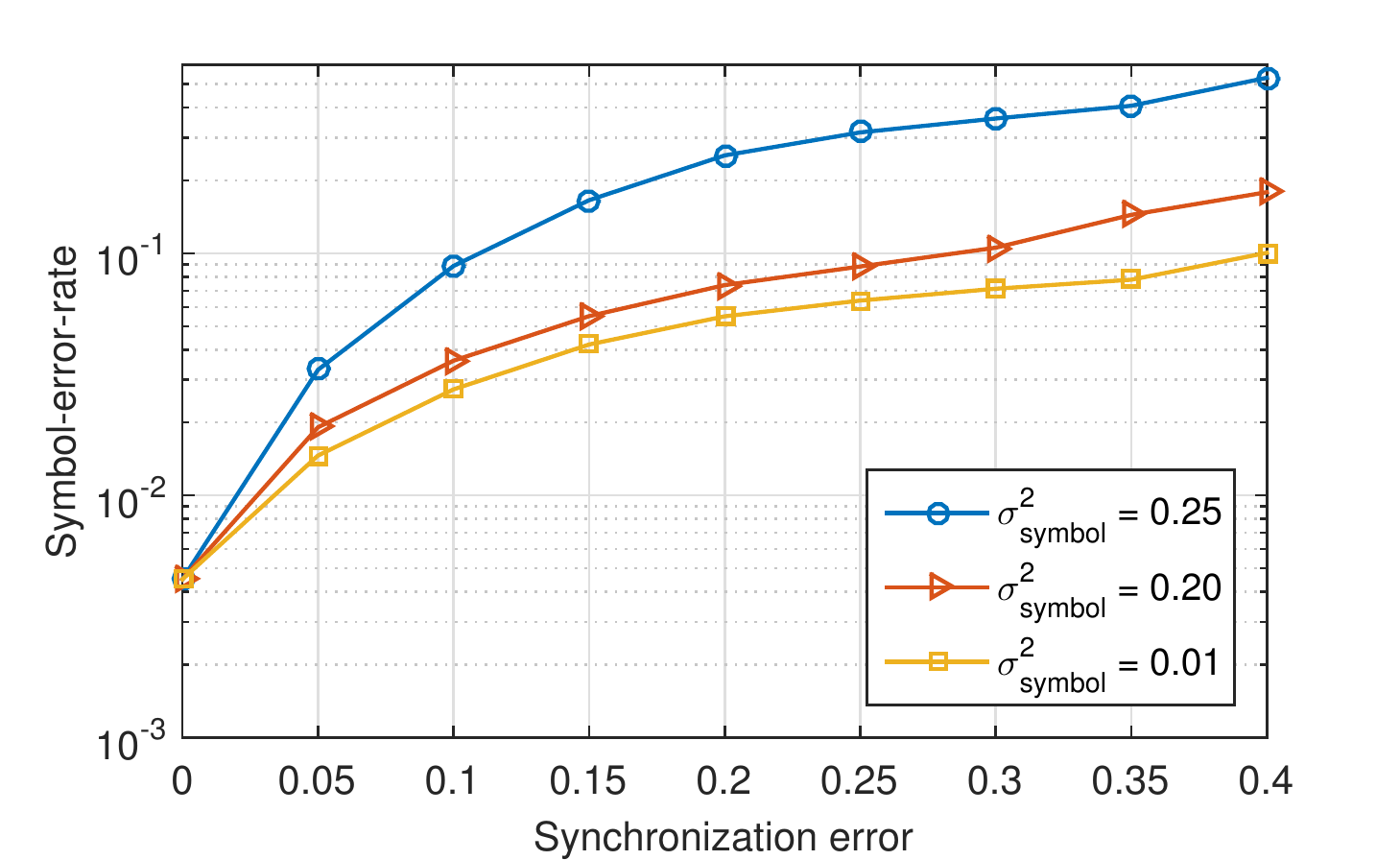}
	\caption{SER performance with the normalized synchronization error $\bar{e}$.}
	\label{fig:SyncError}
     \vspace{-0.14in}
\end{figure}

As shown in Fig.~\ref{fig:SyncError}, the SER increases with the increasing of synchronization error. Since most of the transmitted molecules (both synchronization and information molecules) hit the receiver surface at the beginning of the symbol interval, a small error in the estimation of starting of the symbol duration has adverse impact on the symbol detection based on the received molecule number in MCvD systems. Moreover, in high values of  the standard deviation of the unequal symbol duration, the SER is higher with the increase of synchronization errors. 
%%%%%%%%%%%%%%%%%%%%%%%%

%%%%%%%%%%%%%%%%%%%%%%%%%%%%%%%%%%%%%%%%%%%%%
%%%%%%%%%%%%%%%%%%%%%%%%%%%%%%%%%%%%%%%%%%%%%
%%%%%%%%%%%%%%%%%%%%%%%%%%%%%%%%%%%%%%%%%%%%%
\section{Conclusion}\label{sec:Conclusions}
In this paper, we proposed synchronization scheme for MCvD systems with two types of molecules, namely synchronization and information molecules for the transmission. 
We have exploited the fact that the concentration of received molecules heavily depends on diffusion coefficients of the transmitted molecules. In the proposed system, the synchronization molecules arrive earlier compared to the information bearing molecules, thus the receiver detects the peak of synchronization molecules which act as an indicator for the starting of the information block. In our proposed frameworks, we have considered two cases when both of the synchronization and information molecules are released at the same time and information molecules are released after an offset to the release time of synchronization molecules. First, we analyzed the performance of the proposed system with eye-diagram and continued with BER analysis. From the extensive results, we also evaluated the BER performance of the proposed frameworks with synchronization errors. This work can be extended to see the tradeoff between the cost in term of the number of synchronization molecules and BER performance in the presence of synchronization error. The future work will include to find the optimum offset value in the presence of drift in molecular medium for the MCvD systems.

\section*{Acknowledgments}
This work is partially supported by Guangdong Prov. Key Lab of Petrochemical Equipment Fault Diagnosis, China and the Government of Catalonia's Secretariat for Universities and Research via the Beatriu de Pinos postdoctoral program.

\bibliographystyle{IEEEtran}
\bibliography{MC-Bib}
\end{document}